# As-Li electrides under high pressure: superconductivity, plastic, and superionic states

Zhongyu Wan[1], Wenjun Xu[1], Tianyi Yang[1], Ruiqin Zhang[1][2]*

[1] Department of Physics, City University of Hong Kong, Hong Kong SAR 999077, People's Republic of China

[2] Beijing Computational Science Research Center, Beijing, 100193, People's Republic of China

*E-mail: aprqz@cityu.edu.hk

**Abstract**

Inorganic electrides are a new class of compounds catering to the interest of scientists due to the multiple usages exhibited by interstitial electrons in the lattice. However, the influence of the shape and distribution of interstitial electrons on physical properties and new forms of physical states are still unknown. In this work, crystal structure search algorithms are employed to explore the possibility of forming new electrides in the As-Li system, where interstitial electrons behave as 1D electron chains (1D electride) in *Pmmm* phase of $AsLi_7$ and transform into 0D electron clusters (0D electride) in *P6/mmm* phase at 80 GPa. The *P6/mmm* phase has relatively high superconductivity at 150 GPa ($T_c$=38.4K) than classical electrides, even at moderate pressure with $T_c$=16.6K. The novel superconducting properties are conjectured to be possibly due to three Van Hove singularities at the Fermi level. In addition, a Dirac cone in the band has been observed, expanding the sources of Dirac materials. The survival of $AsLi_7$ at room temperature is confirmed by molecular dynamics simulation at 300 K. At 1000 K, the As atoms in the system act like solid, while a portion of the Li atoms cycle around the As atoms, and another portion of the Li atoms flow freely like liquid, showing the novel physical phenomenon of the coexistence of the plastic and superionic states. This suggests that the superionic and plastic states cannot only be found in hydrides but also in the electride. Our results indicate that superconducting electride $AsLi_7$ with superionic and plastic states can exist in Earth's interior.

Inorganic electrides are a new class of compounds in which some of the surplus electrons in their crystal structure break away from original atomic orbitals and enter into interstices[1], forming aggregated clusters in interstices and acting as interstitial quasi-atoms (ISQ)[2,3]. This has led to various novel properties and applications, in which the inorganic electrides act as catalytic[4], electrode[5], superconducting[6], and insulating materials[7]. In addition, modulating the distribution of interstitial electrons via changing chemical components or external conditions can lead to novel properties of the electride[8,9], such as the superconductivity by doping O atoms in electride $Nb_5Ir_3$[8], and the transformation of the insulating phase $Ca_2N$-II into the metallic $Ca_2N$-I by performing high pressure[10,11]. This suggests that the shape of ISQs also impacts their performance to a large extent[9,12]. Therefore, finding interstitial electrons with different shapes is key to expanding the application of electrides[13].

Unfortunately, their high electron activity makes most electrides extremely susceptible to decomposition under environmental conditions[14,15], resulting in the limited development of atmospheric pressure electrides. Recently, interstitial electrons induced under high-pressure conditions have been proved to form stable lattices, particularly in non-metallic lithium-rich compounds such as Li-P[16,17], Li-C[18] and Li-Te[19] systems. This provides a new approach to the discovery of unknown electrides. At the same time, modulation of electronic structures by external pressure makes them promising as potential superconductors[20]. Element As has a similar electronegativity (2.0 vs 2.1, 2.5, and 2.1) and valence electron number compared to P, C, and Te[21], which suggests that unknown superconducting electrides maybe exist in the As-Li system. Therefore, in this work we perform a crystal structure search for lithium-rich compounds of arsenic at high pressures to find new high-pressure electrides and physical properties with different shapes of interstitial electrons.

The CALYPSO software[22] is used to perform a search for possible $As_xLi_y$ ($x$=2, $y$=2-16) crystal structures at 50, 100, and 150 GPa. The Vienna Ab-initio Simulation Package[23] is used to perform density functional calculations, with the projected-augmented wave pseudopotentials method[24] to describe the electron-ion interaction, using $3d^{10}4s^24p^3$ and $1s^22s^12p^0$ valence electronic structures of As and Li, respectively. Generalized gradient approximation and Perdew-Burke-Ernzerhof as exchange-correlated functional[25] are employed with a cutoff energy of 700 eV and k-point grid density of $0.03 \times 2\pi$ based on the Monkhorst-Pack method[26]. The ionic positions and cell parameters are optimized with the criteria of energy and forces being $10^{-5}$ eV and 0.002 eV/Å, respectively. The ten lowest structures for each stoichiometric ratio are subjected to zero-point energy correction based on the quantum effects[27]. Density functional perturbation theory available in the PHONOPY code[28] is used to calculate the phonon dispersion curve of a 4×4×4 supercell. Ab initio molecular dynamics (AIMD) simulations of a 3×3×2 supercell with 144 atoms is performed for 7000 steps with a step size of 1 fs at the given temperature T = 300 K, 1000 K, and 2000 K with the NVT ensemble, respectively. The QUANTUM ESPRESSO package[29] is used to calculate the electron-phonon coupling (EPC) properties in linear response theory, using the ultra-soft pseudopotential with a cutoff energy of 120 Ry for As and Li. For *Pmmm* and *P6/mmm* based on 16×12×12 and 16×16×12 k-point grids, respectively, the Methfessel-Paxton first-order spread is set to 0.01 Ry, and their irreducible q-points on the grid of 4×3×3 and 4×4×3 centered on Γ are used to first-order perturbation and dynamics matrix calculations.

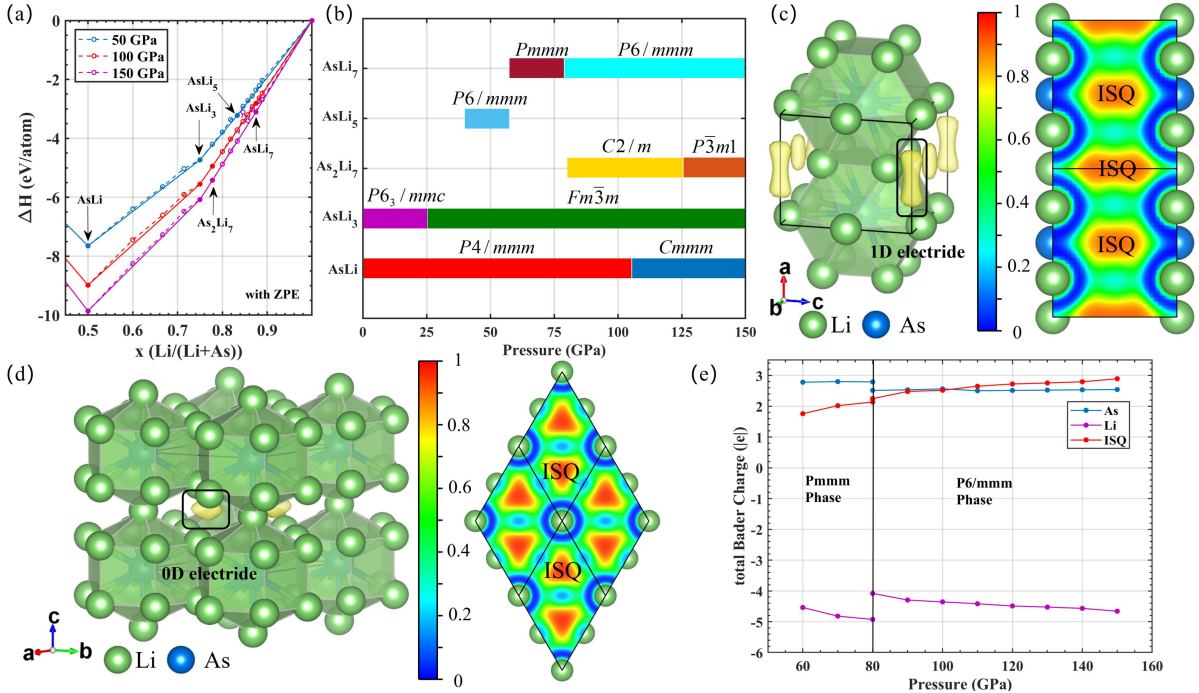

**Figure 1. (a) Convex-hull of As-Li compounds at different pressures. (b) High pressure phase diagram of stable As-Li compounds at 0-150 GPa. (c) ELF of *Pmmm* phase AsLi$_7$ at 80 GPa (isosurface = 0.85). (d) ELF of *P6/mmm* phase AsLi$_7$ at 150 GPa (isosurface = 0.85). And (e) Bader charge of AsLi$_7$ from 60 to 150 GPa.**

Convex-hull lines at 50, 100, and 150 GPa with 0 K is obtained by performing a crystal structure search (Figure 1(a)). Compounds located on the convex-hull do not decompose into corresponding monomers[30,31] and other components and are thermodynamically stable. It can be seen that AsLi, AsLi$_3$, As$_2$Li$_7$, AsLi$_5$, and AsLi$_7$ are present on these lines, and their high-pressure phase diagrams from 0 to 150 GPa are derived by further refined structure searches (Figure 1(b)). The predicted atmospheric pressure structure[32] of $AsLi_3$ proves the credibility of the structure search. The electron localization function (ELF) is suggested[33] to be set to 0.85 to find electrides. However, only $AsLi_7$ exhibits strongly localized interstitial electrons in these compounds. It is stable as the *Pmmm* phase at 60 GPa and transforms to the *P6/mmm* phase at 80 GPa until 150 GPa. Interestingly, the shapes of these interstitial electrons are entirely different. In the *Pmmm* phase (Figure 1(c)), the As atom forms a super-coordinated 18-sided structural unit [As-$Li_{14}$] with 14 Li atoms, sharing a crystal plane between units, while the interstitial electrons are localized in a direction parallel to the a-axis, forming a 1-dimensional electron chain. In the *P6/mmm* phase, the super coordinated structural unit [As-$Li_{14}$] remains constant, but only one crystal axis is shared between units, and the 1-dimensional electron chain is transformed into a 0-dimensional electron cluster.

The phenomenon of super coordination reveals anomalous charge transfer in the electride $AsLi_7$. To obtain the valence states of the different atoms, Bader charges[34] are employed for in-depth analysis (Figure 1(e)). It can be seen that the Bader charge of As remains constant for both the *Pmmm* and *P6/mmm* phases, while the Li atom and ISQ decrease and increases significantly, respectively, implying that Li atoms provide the interstitial electrons. This is because 4p orbital of As is filled with 3 electrons and cannot accommodate any more electrons, while the pressure drives the interatomic distance shorter, leading to greater Coulomb repulsion and minimizing the energy of this system 2s electron of Li which breaks away from the nucleus into interstices.

Given the effect of pressure on electron transfer, it is necessary to study its electronic structure modification and physical properties. From Figures 2 (a) and (b), it can be generalized that pressure can cause the overall density of electron states (DOS) to migrate downwards. It also explains the quasi-atomic role of interstitial electrons, which are very close to As and Li-2s, although the DOS of ISQ at the Fermi level is not as good as Li-2p, suggesting a certain contribution of ISQ to the metallicity of electride $AsLi_7$. And there are far more low-level DOS in As than high-level ones because the entire shell layer structure formed at the valence layer prevents the As orbitals from reaching higher energy. In addition, As and Li atoms have similar trends at DOS, implying strong coupling between atoms and thus promising potential superconductors. The modified McMillan-Allen-Dynes equation[35] calculates the superconducting properties at different pressures:

$$T_c = \frac{\omega_{\log}}{1.2k_B} \exp\left[-\frac{1.04(1+\lambda)}{\lambda - \mu^*(1+0.62\lambda)}\right] \qquad (1)$$

where $\omega_{\log}$ is the log-averaged phonon frequency, $k_B$ the Boltzmann constant, $\mu^*$ the Coulomb pseudopotential ($\mu^*=0.10$) and $\lambda$ the EPC constant. Figure 2(c) shows the variation of superconductivity with pressure, where the *P6/mmm* phase has the highest $\lambda$, the smallest $\omega_{\log}$, and the largest $T_c$ at 150 GPa. The physical properties are often determined by its electronic structures, and in order to obtain the physical insight and mechanism of the highest $T_c$, its band structure is shown in Figure 2(d), which reveals that near the Fermi level, bands are flat in the Γ→M and L→M directions, in line with the "flat-steep band" characteristic[36] of general superconductors. Furthermore, there are three Van Hove singularities[37] (first-order derivatives of band curves is 0) on the Fermi level, located at M point and L→H direction, respectively. Their presence implies that the system has an excellent performance in superconductivity. We also find a Dirac cone[38] at the H point, suggesting that in addition to topological insulator surfaces[38] and Fe-based superconductors[39], it also exists in electrides with potentially high electron mobility and topological properties. Figure 2(e) shows the role of phonon dispersion on superconductivity. The absence of phonon spectra at imaginary frequencies indicates that it is kinetically stable, and the phonon DOS of As atom is more concentrated at lower frequencies than Li atoms due to its larger atomic mass. The calculated Eliashberg spectral function reflects a 66% contribution of phonons below 10 THz to the EPC constant at 150 GPa, implying that the superconductivity of electride $AsLi_7$ is dominated by low-frequency phonons and that the strong electron-phonon coupling makes its superconductivity critical temperature higher than that of known electrides in the range of 0-150 GPa[40,41].

Since our DFT calculations are based on a temperature of 0 K, a condition far away from the actual environment, as can be seen by implementing AIMD at 300 K to obtain root-mean-square displacement curves (RMSD) for different atoms (Figure 3(a)). Whether it is As or Li, their RMSD remains constant, the slope of curves $k \approx 0$, indicating that there is no significant change in the displacement of the atoms, and the trajectories are shown in Figure 3(b), where they act like solids, which means it can survive at room temperature. The temperatures of 1000 and 3000 K are further used to investigate the kinetic behavior of $AsLi_7$ at ultra-high temperatures, and the RMSD curves at 1000 K show slopes $k_{As} \approx 0$ and $k_{Li} > 0$ for different atoms, indicating that the Li atoms are more liquid-like compared to As atoms which maintain a solid character. At temperatures up to 3000 K, there are results for $k_{As}>0$ and $k_{Li}>0$, proving that Li and As atoms behave as liquids and the whole system "melts". The atomic trajectories at 1000 K exhibit strange physical phenomena (Figure 3(c)), with As atoms vibrating near their initial position, while part of the Li atoms moves around the As (plastic state[42]) and part of the Li atoms flow freely (superionic state[43]). This shows that the plastic and superionic state can co-exist in the electride AsLi$_7$ at this temperature. At 3000 K, all atoms become "liquid," confirming the "melting" of the system. The relationship between the "melting points" of different elements is concluded as $T_{Li}<T_{As}$. The energy homogeneity theorem and radial distribution functions (RDF) of As-Li (Figure 3(e)) also justify the simulation results, where the energy at equilibrium is evenly distributed between the degrees of freedom and the displacement of the atoms is stronger due to the lighter mass under the same kinetic energy conditions. Multiple peaks are clearly demonstrated in the RDF at 300 K. However, as the temperature increases, the height of the peaks decreases and becomes broader, indicating a transition from the solid to the amorphous state.

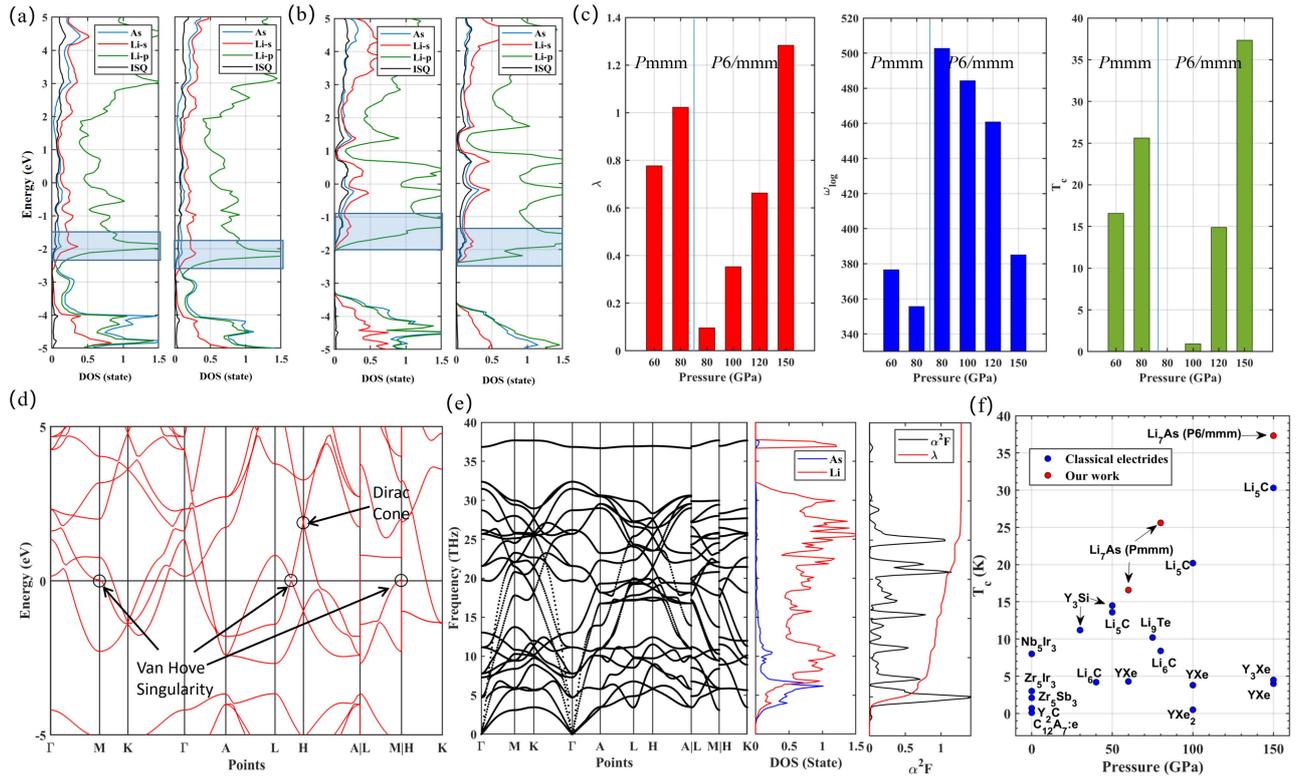

**Figure 2. (a) The density of states of *Pmmm* phase AsLi$_7$ at 60 and 80 GPa. (b) Density of states of *P6/mmm* phase AsLi$_7$ at 80 and 150 GPa. (c) Variation of EPC coefficients, $\omega_{log}$, and T$_c$ with pressure. (d) Electronic structure of the *P6/mmm* phase at 150 GPa. (e) Phonon dispersion curves, phonon density of states, and Eliashberg spectral functions for the *P6/mmm* phase at 150 GPa. And (f) comparison of superconductivity with classical electrides[40,41].**

Since the Earth's interior[44] is also exposed to the same high pressure and temperature conditions and As and Li have high abundances within the Earth, we conjecture that the electron compound AsLi$_7$ can be found in it.

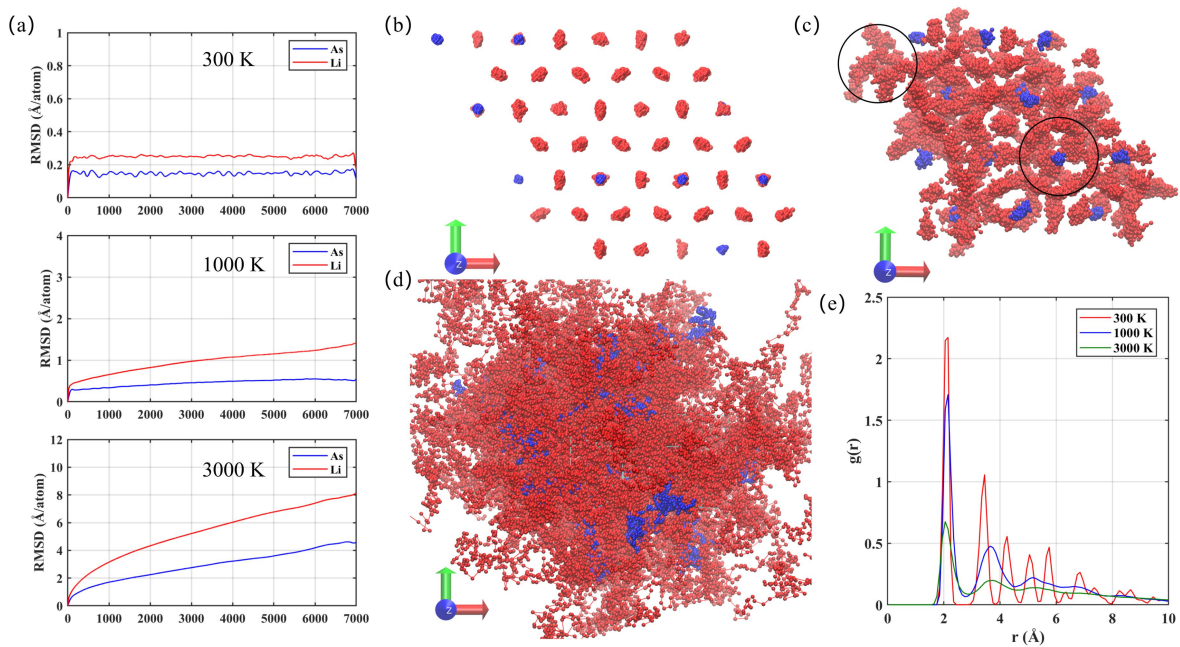

**Figure 3. (a) RMSD curves of As and Li at different temperatures. (b) Trajectories at 300 K (Li in red, As in blue). (c) Trajectories at 1000 K. (d) Trajectories at 3000 K. And (e) RDF curves at different temperatures.**

In this work, the possible phases of lithium-rich compounds of arsenic at high pressure are obtained by a structure search algorithm, and a series of high pressure phase diagrams of As-Li are determined. AsLi$_7$ is found to be capable of forming 1D and 0D electrides in the phase of *Pmmm* and *P6/mmm*, respectively. The Bader charge analysis confirms that the interstitial electrons are contributed by Li atoms. The *P6/mmm* phase has the highest superconductivity at 150 GPa compared to classical electrides, and the presence of three Van Hove singularities at the Fermi level in the band structure is a possible reason for the novel performance. In addition, the Dirac cone at the H point reveals potentially high electron mobility and topological properties, expanding the sources of Dirac materials. Further, AIMD confirms the possibility of existence at room temperature and the appearance of novel coexistence of plastic and superionic states. Since the simulated environment of AIMD is close to that of the Earth's interior, so the electride may exist in it, providing a theoretical guide to the experimental discovery of novel matter.

The authors declare that they have no known competing financial interests or personal relationships that could have appeared to influence the work reported in this paper.

The work described in this paper was supported by grants from the Research Grants Council of Supplementary materials to this article can be found online.